\documentclass{ws-ijgmmp}

\begin{document}

\markboth{D. MOMENI, M. RAZA, R. MYRZAKULOV}
{ANALYTICAL COEXISTENCE OF
$s, p, s + p$ PHASES OF A HOLOGRAPHIC SUPERCONDUCTOR}

%
\catchline{}{}{}{}{}
%

\title{ANALYTICAL COEXISTENCE OF
s, p, s + p PHASES OF A HOLOGRAPHIC SUPERCONDUCTOR
}

\author{DAVOOD MOMENI}

\address{Eurasian International Center for Theoretical Physics
and Department of General \&
Theoretical Physics, Eurasian National University, Astana 010008, Kazakhstan\\
\email{d.momeni@yahoo.com} }

\author{MUHAMMAD RAZA}

\address{Department of Mathematics, COMSATS Institute of Information Technology
(CIIT), Sahiwal Campus, Pakistan\\
mreza06@gmail.com }

\author{RATBAY MYRZAKULOV}

\address{Eurasian International Center for Theoretical Physics
and Department of General \&
Theoretical Physics, Eurasian National University, Astana 010008, Kazakhstan\\
\email{rmyrzakulov@gmail.com} }
\maketitle

\begin{history}
\received{(Day Month Year)}
\revised{(Day Month Year)}
\end{history}

\begin{abstract}
We analytically study\ the critical phase of a mixed system of the
$U(2)$ gauge fields and global symmetry on the boundary using
gauge/gravity. A variational minimization problem has been
formulated. The numerical results pertinently show that there exists
a minimum chemical potential in which both scalar (s-wave) and
vector (p-wave) condensates exist in a mixture, as well as in the
distinct s- and p-phases. This result is obtained by breaking the
symmetry into $U(1)$ symmetry and rotational symmetry. While the
analytical solutions of condensates and charge densities are
achieved in both cases: the balanced and unbalanced holographic
superconductors. This is the first analytical study of the
coexistence of two modes of the superconductivity with different
order parameters. The realistic model consists of two different
phases of the superfluidity in Helium.
\end{abstract}


\section{Introduction}

Anti de-Sitter/ conformal field theory (AdS/CFT) proposed by
Maldacena conjecture states that any weakly gravitational model in
the bulk has a dual boundary quantum field theoretic description
through a set of the dual quantum relevant operators $\mathcal{O}$
\cite{maldacena}. Especially a high temperature superconductor, a
strongly coupled system, has a good well defined gravitational dual
using this conjecture \cite{HSC1,HSC2,HSC3}. The conjecture has been named as
the gauge/gravity duality which has been studied in several papers.
Depending upon the kind of the condensates used, we have scalar
(s-wave), vector broken rotational symmetry (p-wave) or d-wave and
more. From the point of view of symmetry, spontaneously breaking of
different kinds of the holographic superconductors corresponds to
different schemes of the breaking of different symmetries. In brief,
regarding the scalar condensates, the symmetry is broken to $U(1)$.
Further, p-wave is the result of rotational symmetry. But a mixed
phase $s+p$ is due to the breaking of the two $U(1)$ symmetries to two
distinct s- and p-phases. It
comes from the symmetry breaking cascade, that is of the form $%
U(2)\rightarrow U(1)\times U(1)$. Bulk action can be included by
higher order curvature corrections like Gauss-Bonnet, Weyl and etc.
The role of these higher order corrections and their dual
descriptions on the boundary using AdS/CFT has been widely studied
in the literature.

Very recently, the problem of the existence or coexistence of two
modes s- and p-waves simultaneously has been investigated in the
literature. It has been shown that there is a possibility to have
two s-wave phases with two different scalar condensates
\cite{Musso:2013ija}. This study also extended to include the case
of the spontaneously symmetry breaking under the rotational symmetry
by a vector as a possibility for the coexistence of two p-wave order
parameters \cite{Amoretti:2013oia}. But there is an
intermediate phase between s- and p-phases. This phase is known as the $s+p-$%
phase. In this phase we have simultaneously two kinds of the
relevant operators on the boundary. The CFT description of it has a
direct interpretation for $s+p$ as a candidate of a mixed phase.
This model is based on the $U(2)$ gauge fields and numerically studies
the $s+p$ wave phase in a typical holographic
superconductor \cite{Nie:2013sda}. Further, the competition between
the conventional s-wave and the triplet Balian-Werthamer or the
B-phase pairings in the doped three dimensional narrow gap
semiconductors has been investigated \cite{cmt}.

Firstly, the existence or coexistence problem was related to the
usage of
two vector order parameters and was explained in terms of the AdS/CFT in Ref.%
\cite{Zayas:2011dw}. As it was shown that, in this mixed system, the
phase transition can be controlled by the form of the interaction.

The model used in this work is based on a toy model of the multiband
holographic superconductors proposed originally in
\cite{Krikun:2012yj} and extended in details in
\cite{Amado:2013xya}. This multiband model was used to describe the
AdS/CFT description of two-component superfluidity and it has been
shown that the system under a perturbation becomes unstable.
Actually, the system under such kind of the instability moves towards the $%
s+p$ wave phase -an intermediate phase, in which the free energy of
the system minimized and this intermediate phase is preferred than
the other pure phases.

So far it is believed that under some restrictions, upper than a
critical chemical potential $\mu _{c}$, there exists a possibility
to have both s- and p-phases in a mixture, it is called as the
$s+p$-phase. It is a transitive mode and in a very highly condensed
regime of the dual system on the boundary it undergoes to the purely
p-wave phase. In this paper, we study some basic analytic properties
of such mixed $s+p$ phase using a highly motivated semi analytical
method of functional theory. We show that the system of highly
non-linear coupled differential equations can be transformed into a
self-adjoint  system of auxiliary functions $F$ and $G$. By using
the variational methods we obtain the minimum chemical potential of
the different modes of the superfluidity. In the case of $s+p$-phase
we used analytic expressions for condensates which remain valid only
near the critical point. Also, the non-linear relation between the
charge densities and the chemical potentials near the critical point
has been studied. We show that the linear approximation, so far, is
a valid approximation. A qualitative comparison between the values
of free energy shows the best results for the mixed $s+p-$phase.

\section{On Gauge/Gravity Picture of the Holographic Superconductors}

Consider a weak gravitational bulk action in four dimensions taken
as the large color number limit of a super string theory. In this
limit the gravity sector is decoupled and behaves like a classical
matter field in the spacetime manifold. We assume that the metric
has a static and spherically symmetric tensor form. Asymptotically
the metric must be Anti de-Sitter (AdS). If we calculate the action
of the bulk we will be able to compute the point functions (two,
...) as local functions of the fields. Here by fields we mean scalar
fields mainly. With action we are able to compute the partition
function $Z$ of the bulk. The keynote here is that the asymptotic
behavior of the two points function of the scalar fields can be read
off in terms of the source of some relevant operators of a typical
conformal field theory (a field theory of operators with traceless
energy momentum tensor and with a definite central charge) on the
boundary. Mathematically, the system in the bulk (for example a
black hole with non zero temperature on horizon or a soliton with
zero temperature) remains in the thermal equilibrium with boundary
under the following necessary and sufficient conditions:
\[
Z_{Bulk}=Z_{CFT}\Rightarrow T_{BH}=T_{CFT}.
\]%
The above brief introductory section describing the relation between
the correlation functions of some scalar matter fields on a bulk and
the expectation values (the source terms) of some quantum operators
(strictly speaking some relevant quantum operators) is considered as
the gauge/gravity duality in its very simple interpretation. A
technical discussion must be done to cover all vague points. For
example, which kind of the duality exists if the gravity deals with
some higher order corrections like Gauss Bonnet terms or Weyl ones.
This duality is between a bulk and a boundary with planar horizon. We need to review some mathematical steps ensuing
this duality in this section.

The conformal field theory is a metric conformally invariance theory
on the boundary. Maldacena gauge/gravity interpretation is related
to such quantum theory on boundary and a higher dimensional gravity
model. To have a CFT, we need that the total Hamiltonian of a
typical quantum theory remains invariant under a certain non
singular metric conformal transformations as the following:
\[
g_{\mu \nu }\rightarrow \Omega (x^{\alpha })g_{\mu \nu }.
\]%
We consider a typical general coordinate transformation like $x^{\mu
}\rightarrow x^{\mu }+a^{\mu }$. Here $a^{\mu }$ is  a translation in the spacetime coordinate. Under this local
transformation, we assume that it is possible to write down a
conserved current by contraction of the $a^{\mu } $ with the energy
momentum (EM) tensor of this field theory $T^{\mu \nu }$ like:
\[
J^{\beta }=T^{\alpha \beta }a_{\alpha }.
\]%
By EM tensor we mean the effective EM tensor of the CFT. We apply
the continuity equation on the current and since all quantities are
local quantities so the continuity equation is local and can be
written as:
\[
\partial _{\beta }J^{\beta }=0.
\]%
By substituting the current we find that EM tensor must be traceless
and in the following form $T_{\beta \mu }g^{\beta \mu }=0$. The
strategy of using the CFT is also useful to compute the degrees of
freedom of the fields or central charge as well. Back to the
conformal transformation, we have three types of the generators for
three types of the possible symmetries. The last one corresponding
to the scaling symmetry is called the dilatation operator. The
operation of this quantum operator is understood by applying it to a
unitary operator of CFT as $\mathcal{O}$ satisfying the following
operator equation:
\[
D\mathcal{O}=\Delta \mathcal{O}.
\]%
This equation is in the form of an eigen value function operator.
The eigen value $\Delta $ is called the conformal dimension and as
it has been shown, its value is defined by the mass of the scalar
fields on the boundary. In the four dimensional bulk theory where
the boundary is three dimensional, this preliminary treatment of the
$AdS_{4}/CFT_{3}$, gives us:
\[
\Delta (\Delta -3)=m^{2}L^{2}.
\]%
Here $m$ is the mass of the scalar fields in the bulk and $L^{2}=-\frac{6}{%
\Lambda }$ is the AdS radius of the spacetime(commonly we set L=1).
For vector fields we need to replace $\Delta (\Delta -3)\rightarrow
\Delta (\Delta -1)$.

To study the condensed matter physics we need scaling invariant
quantum theories. From the point of view of the renormalization
group, such theories can be addressed as the ending points of the
flows. To apply a scaling invariant theory it is required to keep
the operators under the scaling invariance as the following:
\[
\mathcal{O}=\lambda ^{D}\mathcal{O}(\lambda x).
\]%
Three kinds of the quantum operators live on the boundary treating
the CFT:
relevant, irrelevant and marginal. In order to write the CFT Hamiltonian ($%
H_{CFT})$ and to avoid a non renormalizable quantum theory we only
use the relevant operators. So, the total action including the free
part and the interaction action reads:
\[
S=S_{free}+\int {d^{4}x\Sigma _{i}g_{i}\mathcal{O}_{i}}.
\]%
The first term denotes the free fields contribution, the last part
is the interaction between the fields and $g_{i}$ are the
interaction coupling constants. If the model is renormalizable then
the interaction couplings should satisfy the following first order
(always) integrable system:
\[
\frac{dg_{i}}{d\log {\epsilon }}=\beta _{i}(g_{j}).
\]%
The integral curves are the solutions (first integrals) of the
renormalization group flow equations. This dynamical system has a
set of fixed points located at $\beta _{i}(g_{j}^{\ast })=0$. Near
these fixed points the quantum theory is scale invariant and we are
able to explain the evolutionary dynamics of the system using a
quantum field theory with invariant structure under scaling
transformations. Any small perturbation of this kind of CFT induces
an extra flow. With this statement we can say that gauge/gravity is
a duality and it represents a one to one map between any gauge
theory described by CFT on the boundary and a bulk matter action in
thermal equilibrium with the boundary.

To have a better view on AdS/CFT we start with a simple model of a
single scalar field $\phi $ with the following action:
\[
S\sim \int
{d^{4}x\sqrt{-g}\Big(\mathcal{L}_{CFT}+\mathcal{L}_{m}\Big)}
\]%
Here we set the AdS radius $L=1$. The first term corresponds to the
CFT and the second term induces the quantum operators on the
boundary. Any small deformation from this CFT action, produces a
flow. After this emergence of the flow, the full Hamiltonian of the
system including the quantum operators (induced on the boundary by
the matter action) can be written as the below:
\[
\hat{H}=H_{CFT}+\Sigma _{i}\lambda _{i}\mathcal{O}_{i}.
\]%
Here the summation is taken over all the relevant or irrelevant
operators.
By the definition, a holographic superconductor is a high temperature  superconductor which has a dual gravitational model.
The core of the gauge/duality picture of the holographic
superconductor is that the quantum vacuum expectation value of a
relevant operator $\mathcal{O}_{r}$
has the following form versus temperature and the critical temperature $%
T_{c} $:
\[
<\mathcal{O}_{i}>\sim \sqrt{1-\frac{T}{T_{c}}}.
\]

The main purpose of this dualization is to find the basic properties
like critical exponent, critical temperature and critical chemical
potentials. Since the system of the bulk field equations is highly
non linear, so the common method to solve these equations is
numerical methods based on the shooting tool. It is desired to find
the temperature dependence of the vacuum expectation values of the
boundary operators $<\mathcal{O}_{\pm }>$ or the chemical potentials
$\mu _{c}$ at the critical points at which a phase transition
occurs.

\section{Model}

As we mentioned before our aim here is to study the analytical
properties of a mixed phase $s+p$ of a holographic superconductor.
We start by writing the action of a $3+1$ dimensional bulk with AdS
boundary, by including a $U(2)$ gauge field and a scalar doublet
field, instead of the usual scalar field \cite{Amado:2013lia}:
\begin{equation}
S=\int d^{4}x\sqrt{-g}(\frac{R-2\Lambda }{2\kappa ^{2}})+\int d^{4}x\sqrt{-g}%
\Big(\frac{-1}{4}F_{c}^{\mu \nu }F_{\mu \nu }^{c}-m^{2}|\Psi
|^{2}-|D^{\mu }\Psi |^{2}\Big).  \label{action}
\end{equation}%
The mass of the scalar-doublet remains upper than the Breitenlohner
and Freedman (BF) bound \cite{BF,BF2}. It is believed that this action
explains the mixture of both $s-$and $p-$ phases, and read off the
vacuum field expectation values on the boundary conformal fields
using AdS/CFT. Here the
notation is the same as the usual p-wave holographic superconductors, like $%
D_{\mu }=\partial _{\mu }-iqA_{\mu }$, whereas for the simplicity
and in the probe limit we neglect the back reaction of all matter
fields, we set $q=1$. Also the gauge field is expanded on the basis
of the units (generators) of SU(2) group and Pauli's matrices
$\sigma _{i}$. The only difference is that here we have a normalized
doublet scalar $\Psi $:
\[
\Psi =\left(
\begin{array}{c}
\frac{w(r)}{\sqrt{2}} \\
\frac{\psi (r)}{\sqrt{2}}\
\end{array}%
\right) .
\]

To preserve the symmetry of the AdS metric, we assume that all the
functions
are well defined and are functions(single valued) of the radial coordinate $%
r $. It enables us to perform a radial quantization for an AdS/CFT
construction of the dual model. It is a vital and fundamental
feature of our gauge/gravity approach. We assume that the bulk is an
AdS spacetime in the following form:
\begin{equation}
g_{\mu \nu }=diag[-f(r),f(r)^{-1},r^{2}\Sigma _{2}],\ \ \Sigma
_{2}=diag(1,\sin ^{2}\theta ).  \label{metric}
\end{equation}%
The metric represents a black hole bounced in an AdS and with planar
topology of the horizon. By using the same assumptions on the fields
and following the \cite{Amado:2013lia}, we can write the next set of
the field equations for our system in the probe limit, ignoring the
back reactions (in units $2\kappa ^{2}=1$):
\begin{eqnarray}
\psi ^{\prime \prime }+\Big(\frac{f^{\prime
}}{f}+\frac{2}{r}\Big)\psi
^{\prime }+\Big(\frac{(\Phi -\Theta )^{2}}{4f^{2}}-\frac{m^{2}}{f}-\frac{%
w^{2}}{4r^{2}f}\Big)\psi &=&0  \label{eq1} \\
\Phi ^{\prime \prime }+\frac{2\Phi ^{\prime }}{r}-\frac{\psi
^{2}}{f}(\Phi
-\Theta ) &=&0  \label{eq2} \\
\Theta ^{\prime \prime }+\frac{2}{r}\Theta ^{\prime }+\frac{\psi ^{2}}{f}%
(\Phi -\Theta )-\frac{w^{2}}{r^{2}f}\Theta &=&0  \label{eq3} \\
w^{\prime \prime }+\frac{f^{\prime }}{f}w^{\prime }+\frac{\Theta ^{2}}{f^{2}}%
w-\frac{\psi ^{2}}{f}w &=&0.  \label{eq4}
\end{eqnarray}%
Here prime denotes the derivative with respect to the radial
coordinate $r$ and $\Phi$ and $\Theta$ are the components from the gauge fields of $U(1)$ and $SU(2)$. Previously full numerical solutions for condensates
(scalar and vector) were presented. It has been shown that the mixed
phase $s+p$ exists as a sub dominant phase under $s-$wave for both
the cases of balanced and unbalanced
systems. Our aim here is to study the phase transitions in system (\ref{eq1}-%
\ref{eq4}) using the analytical method, proposed in \cite{siopsis}.
In this method, the keynote is to write down the equations near the
criticality as a variational self-adjoint problem with appropriate
boundary conditions. The eigenvalue of the variational problem
denotes the chemical potential in the dual CFT description. To
minimize the chemical potential, we minimize the functional integral
using an arbitrary trial function. The minimum chemical potential is
obtained by finding the optimum values of the variational
parameters. If we increase the number of the variational parameters,
the values of our semi analytical estimation becomes close to the
numerical results, which can be obtained by solving the system of
the coupled non linear differential equations using shooting method
\cite{Amado:2013lia}.

The normal phase, when the temperature of the system $T>T_{c}$, is
defined
by the following solutions of system of the field equations (\ref{eq1}-\ref%
{eq4}) :
\[
f=r^{2}-\frac{r_{+}^{3}}{{r}},\psi =w=0,\Theta =\mu ^{\prime }(1-\frac{1}{r}%
),\ \ \Phi =\mu (1-\frac{1}{r}).
\]%
Here $\mu $ is the dual chemical potential for $U(1)$ but $\mu
^{\prime }$ comes from the $SU(2)$. They have the same description
using group theory of the $U(2)$ group. If we set $\mu ^{\prime }=0$,
the model is a balanced holographic superconductor. In the
unbalanced case, $\mu ^{\prime }\neq 0$. We always keep this as last
one. The only vital change is to perform a recomputation of free
energy density using the on-shell action, whose balanced case does
not contain the constant term $-\frac{1}{2}\mu ^{\prime
}\rho ^{\prime }$. We are able to modify the equations given by (\ref{eq1}-%
\ref{eq4}) with a non unit $r_{+}$, but to compare our results with
the numerical ones in \cite{Amado:2013lia}, we set $r_{+}=1$.

We have two kinds of the condensates, one of these is the leading
term of the scalar field $\psi $ on the AdS boundary:
\begin{equation}
\psi \cong \frac{<\mathcal{O}_{+}>}{r^{\Delta _{+}}}+\frac{<\mathcal{O}_{-}>%
}{r^{\Delta _{-}}},
\end{equation}%
where, as usual, $\Delta _{\pm }$ denotes the conformal dimension,
for renormalizability we take $\Delta _{\pm }\geq 1$. It corresponds
to the
choice of the scalar doublet mass $m^{2}=-2$, that is upper to the BF bound $%
m_{BF}^{2}=-\frac{9}{2}$. As we know $\mathcal{O}_{\pm }$ denotes
the scalar order parameter. Here, we also have a current operator
$J_{x}$, which is related (as the source term) to the AdS asymptotic
solution of field $w$ as the following:
\begin{equation}
w\cong \frac{<J_{x}^{1}>}{r}+<J_{x}^{0}>.
\end{equation}%
We need to fix our quantization scheme. We choose to switch-off$<\mathcal{O}%
_{-}>=0,<J_{x}^{0}>=0,<J_{x}^{1}>=<J_{x}>$, so that the set of the
asymptotic solutions in the AdS boundary read:
\[
\Phi =\mu -\frac{\rho }{r},\ \Theta =\mu ^{\prime }-\frac{\rho ^{\prime }}{r}%
,\ \ w=\frac{<J_{x}>}{r},\ \ \psi =\frac{<\mathcal{O}_{+}>}{r^{2}}.
\]%
Our main goal is to find the explicit expressions with respect to
the chemical potentials of the $\{<J_{x}>,<\mathcal{O}_{+}>\}$ (we
choose $x$ as the preferred direction). We have two regimes: the
balanced case of the superconductors with $\mu ^{\prime }=0,$ and
for the unbalanced case, we take $\mu ^{\prime }\neq 0$. We keep
$\mu ^{\prime }$ as a non zero parameter. The case of the unbalanced
superconductors will be studied as a limiting case $\mu ^{\prime
}\longrightarrow 0$.

In terms of the new coordinate $z=\frac{1}{r}$, the field equations (\ref%
{eq1}-\ref{eq4}) for $m^{2}=-2$ have been written in the following
forms:
\begin{eqnarray}
\psi ^{\prime \prime }-\frac{(2+z^{3})}{z(1-z^{3})}\psi ^{\prime }+\Big[%
\frac{(\Phi -\Theta )^{2}}{4(1-z^{3})^{2}}+\frac{2}{z^{2}(1-z^{3})}-\frac{%
w^{2}}{4(1-z^{3})^{2}}\Big]\psi &=&0  \label{eq11} \\
\Phi ^{\prime \prime }-\frac{\psi ^{2}}{z^{2}(1-z^{3})}(\Phi -\Theta
) &=&0
\label{eq22} \\
\Theta ^{\prime \prime }+\frac{\psi ^{2}(\Phi -\Theta )}{z^{2}(1-z^{3})}-%
\frac{w^{2}\Theta }{1-z^{3}} &=&0  \label{eq33} \\
w^{\prime \prime }-\frac{3z^{2}}{1-z^{3}}w^{\prime }+\Big[\frac{\Theta ^{2}}{%
(1-z^{3})^{2}}-\frac{\psi ^{2}}{z^{2}(1-z^{3})}\Big]w &=&0.
\label{eq44}
\end{eqnarray}%
Our aim is to solve (\ref{eq11}-\ref{eq44}) semi analytically.

\section{Variational Method}

Based on the study in \cite{siopsis}, we can find the minimum of the
$\mu $ and $\mu ^{\prime }$ using a variational method. The starting
point is that we make the following substitutions in the fields
which are valid only in vicinity of the critical point:
\begin{eqnarray}
\Phi &\approx &\mu (1-z),\ \ \Theta \approx \mu ^{\prime }(1-z), \\
w &\approx &<J_{x}>zF(z),\ \ \psi \approx D_{+}z^{2}G(z),D_{+}\equiv <%
\mathcal{O}_{+}>.
\end{eqnarray}%
We apply these functions in the equations (\ref{eq11}, \ref{eq44}). Here $%
F,G $ are two trial functions to minimize the variational integral:
\begin{eqnarray}
F^{\prime \prime }+\frac{(2-5z^{3})}{z(1-z^{3})}F^{\prime }+\Big[-\frac{3z}{%
1-z^{3}}+\frac{\mu ^{\prime 2}}{(1+z+z^{2})^{2}}-\frac{D_{+}^{2}z^{2}G^{2}}{%
1-z^{3}}\Big]F &=&0  \label{F} \\
G^{\prime \prime }+\frac{(2-5z^{3})}{z(1-z^{3})}G^{\prime }+\Big[-\frac{4z}{%
1-z^{3}}+\frac{\left( \mu -\mu ^{\prime }\right) ^{2}}{4(1+z+z^{2})^{2}}-%
\frac{<J_{x}>^{2}z^{2}F^{2}}{4(1-z^{3})^{2}}\Big]G &=&0.  \label{G}
\end{eqnarray}%
We call these equations as `F' and `G' respectively. Now we write
the system (\ref{F}, \ref{G}) in the Sturm-Liouville (SL) form as
the following:
\begin{eqnarray}
(z^{2}(1-z^{3})F^{\prime })^{\prime }+\Big[-3z^{3}+\frac{\mu
^{\prime
2}z^{2}(1-z)}{1+z+z^{2}}-D_{+}^{2}z^{4}G^{2}\Big]F &=&0  \label{F2} \\
(z^{2}(1-z^{3})G^{\prime })^{\prime
}+\Big[-4z^{3}+\frac{z^{2}(1-z)\left(
\mu -\mu ^{\prime }\right) ^{2}}{4(1+z+z^{2})}-\frac{<J_{x}>^{2}z^{2}F^{2}}{%
4(1-z^{3})}\Big]G &=&0.  \label{G2}
\end{eqnarray}%
Any (SL) system in the form of $(Py^{\prime })^{\prime }+(\lambda
W+Q)y=0$, can be obtained from the Lagrangian density:
\begin{equation}
L=Py^{\prime 2}-(\lambda W+Q)y^{2}.  \label{L}
\end{equation}

From (\ref{L}) it shows that the minimum of the eigen value $\lambda
$ reads as the following:
\begin{equation}
\lambda _{min}=Min\{\frac{\int_{0}^{1}{dz(Py^{\prime 2}-Qy^{2})}}{%
\int_{0}^{1}{dzWy^{2}}}\}.  \label{integral}
\end{equation}%
For system (\ref{F2}, \ref{G2}) we have:
\begin{equation}
P_{F}=z^{2}(1-z^{3}),\lambda _{1}=\mu ^{\prime 2},W_{F}=\frac{z^{2}(1-z)}{%
(1+z+z^{2})},Q_{F}=-D_{+}^{2}z^{4}G^{2}-3z^{3}.
\end{equation}

and for the G equation:
\begin{equation}
P_{G}=z^{2}(1-z^{3}),\lambda _{2}=(\mu -\mu ^{\prime })^{2},W_{G}=\frac{%
z^2(1-z)}{4(1+z+z^{2})},Q_{G}=-\frac{<J_{x}>^{2}z^{2}F^{2}}{4(1-z^{3})}%
-4z^{3}.
\end{equation}

Now, we will evaluate the integral (\ref{integral}) for $\lambda
_{1},\lambda _{2}$, with trial functions
\begin{equation}
F(z)\equiv y_{F}=1-bz^{2},G(z)\equiv y_{G},
\end{equation}%
The first trial function $y_{F}$ is chosen. To avoid the singularity
in the $F$ equation it is required to fix $y_{F}$ with auxiliary
boundary conditions as the following:
\[
y_{F}(0)=1,\ \ y_{F}^{\prime }(0)=0.
\]%
But about $y_{G}$ the situation is a little bit different. As we
observe from the $G$ equation, it has an essential pole located at
$z=1$. The pole is second order one. So, it is required that the
following expression remains finite:
\[
\frac{G(z)z^{2}F(z)^{2}}{(1-z^{3})^{2}}<\infty .
\]%
Also in the variational integral we need $y_{G}$ which should be
able to remove the singularity safely. So, the best candidates may
be written as the following trial function:
\[
y_{G}=a+bz+cz^{2},\ \ y_{G}(1)=0,\ \ y_{G}^{\prime }(0)=0.
\]%
By imposing the above boundary conditions, we find $c=-a,b=0$, so
the trial function for G is written as the following
\[
y_{G}=a(1-z^{2}).
\]%
There are so many other acceptable trial functions, but to keep the
simplicity of the results, we choose only these two forms .

We discuss here the integrals involving the variational technique
from the
purely mathematical point of view. Look at the forms of the $Q_{F}$ and $%
Q_{G}$. Since near the critical point both $D_{+},$ $<J_{x}>$ treat
like the small parameters, it is better to define the smallness\ of
the parameters \cite{Kanno:2011cs},
\begin{equation}
\epsilon _{1}=D_{+}^{2},\epsilon _{2}=<J_{x}>^{2}.
\end{equation}%
In these forms the dominators of the integral read as the
following:
\begin{equation}
I_{Q}^{i}=\int_{0}^{1}{dzQ_{i}y^{2}}=\int_{0}^{1}{dzQ_{i}^{0}y^{2}}+\epsilon
_{i}\delta I_{Q}^{i},\ \ i=\{F,G\}.
\end{equation}%
Here,
\begin{eqnarray}
Q_{F}^{0} &=&-3z^{3},\ \ \delta I_{Q}^{F}=\int_{0}^{1}{%
dzz^{4}y_{G}^{2}y_{F}^{2}}, \\
Q_{G}^{0} &=&-4z^{3},\ \ \delta I_{Q}^{G}=\int_{0}^{1}{dz\frac{%
z^{2}y_{F}^{2}y_{G}^{2}}{4(1-z^{3})}}.
\end{eqnarray}%
The integral $\delta I_{Q}^{G}$ diverges at the boundary point
$z=1$, except than:

\begin{equation}
y_{G}|_{z\rightarrow 1}\approx a(1-z^{2})+O((1-z^{2})^{2}).
\end{equation}%
We observe that the above form of $y_{G}$ supports the avoidance of
divergency.

Now, we minimize the integral with respect to the $\{a,b\}$. This
evaluation for both the integrals has been done in below:
\begin{equation}
\mu ^{\prime 2}=Min\{\frac{D_{+}^{2}\left(
0.37296a^{2}b^{2}-1.38528a^{2}b+1.52381a^{2}\right) +40.5b^{2}-60b+45}{%
0.455612b^{2}-1.83347b+2.62765}\}.  \label{mu2}
\end{equation}%
And for another case:
\[
\left( \mu -\mu ^{\prime }\right) ^{2}=Min\{0.114305\left( \left(
6.18929b^{2}-19.8343b+18.3936\right) <J_{x}>^{2}+784\right) \}.
\]

From these last equations, we are able to complete the analysis of
the different phases. We mention here that the transition phases
from s-wave to p-wave in the unbalanced cases are  first order
phase transitions. We mention here that near the critical point,
$D_{+}=D_{+}(\mu ,\mu ^{\prime }|\mu _{c},\mu _{c}^{\prime
}),<J_{x}>=<J_{x}(\mu ,\mu ^{\prime }|\mu _{c},\mu _{c}^{\prime
})>$. So the correct minimization must be following a simultaneously
minimization scheme applied on these two functions. In next section,
when we study the critical behavior and the critical exponent of
these expectation values analytically, we will deduce some
expressions for these. Generally in this case of the superconductors
and due to the mixing of homogeneous and inhomogeneous
superconductors we have the following expressions, valid only near
the critical point $(\mu _{c},\mu _{c}^{\prime })$:
\[
D_{+}\sim (\mu -\mu _{c})^{1/2},\ \ <J_{x}>\sim \Sigma _{i,j}\gamma
_{i,j}(\mu -\mu _{c})^{i}(\mu ^{\prime }-\mu _{c}^{\prime })^{j}.
\]%
Here,
\[
\gamma _{i,j}=\frac{\partial ^{2}<J_{x}>}{\partial (\mu -\mu
_{c})\partial (\mu ^{\prime }-\mu _{c}^{\prime })}|_{(0,0)}.
\]%
Whereas we have ignored the proportionality constants in these
definitions.

\section{Balanced Superconductors}

In this case, we set $\mu ^{\prime }=0$. Our aim here is to show
that there
exists any lower bound $\mu ^{\ast }$ such that for at least two values of $%
\mu \geq \mu ^{\ast }$ we can have two separate regimes of
holographic superconductors, as  purely s-wave (with non zero
$\psi $ but $w=0$) or s+p (with non zero $\psi ,w$) or not.

\textbf{Normal phase}: The normal phase of the system when $T>T_c$
exists as
an exact solution to the field equations $\psi=w=\Theta=0$ and $%
\Phi=\mu(1-z) $. The system under a small perturbation undergoes to
an excited superconducting phase.

\textbf{$s$-wave phase}: In this case with $w=0$, from minimization
of $G$ equation with $<J>=0$, we find that for
\begin{equation}
\mu _{\min }^{s-wave}=9.46653.
\end{equation}%
The numerical result as it has been reported in \cite{Amado:2013lia}
is $\mu _{\min }\approx 8.127$. As we observe that our analytical
estimation is in the same order and in a very good agreement with
the numeric,, although there is some $\%14$
 error in certain circumstances, which is acceptable from the analytical study. . It realizes a two component s-wave superfluidity.

\section{Unbalanced Superconductors}

A simple holographic model of a s-wave unbalanced superconductors
has been proposed \cite{Bigazzi:2011ak}. Before introducing the
analytical approach to the holographic model we review the basics of
an unbalanced superconductor in condensed matter physics. In
unbalanced superconductors the phase transition happens from a
normal superconducting phase to another superconducting phase having
a non zero intrinsic(mechanical) angular momentum. This phase
transition is of the first order but in the balanced one is of the
second order. The final superconducting phase with angular momentum
is called as the LOFF (Larkin-Orchinnikov-Fulde-Ferrel) phase. This
phase transition happens because the populations of two groups of
the fermions are inequal. So that the second phase has a non zero
(inequal) chemical potential. From condensed matter point of view,
this phase transition occurs in some strongly coupled systems and
specially in the quark systems in QCD. So, its experimental
verification in the laboratory is
possible. In the balanced superconductors, initially the system when $%
T>T_{c} $ is located at the normal phase-the ground state phase.
After that in $T<T_{c}$ the system fall down to the superconducting
phase as an excited phase due to the energy considerations
thermodynamically. But in the unbalanced superconductors the first
order phase transition is between two superconducting phases: from
usual superconducting with zero angular momentum to the non zero one
(LOFF) configuration. In holographic picture we need an additional
dual pair of the $(\mu ^{\prime },\rho ^{\prime })$ corresponding to
the new gauge scalar field $\Theta $ in the bulk. Although this pair
comes from another dual field but affects the condensation. From
another view, the phase structure on an unbalanced superconductor
has three different phases. One is normal phase $T>T_{c}$. The
superconductor phase has two distinct phases of the homogenous
superconductor in which the BCS theory basically is valid. But the
tail of the superconductivity looks like a small island in which the
superfluidity is LOFF kind. To cross the border of the BCS
superconducting country to this LOFF tribe, the phase transition is
of first order but the situation is different from normal phase to
the BCS superconductivity in which the phase transition is of second
order. But at the border of the LOFF to the normal phase, system has
another second order phase transition as well as the usual
superconductors. This picture remains the same in any weakly coupled
superconductor.

Following the model proposed in last section to have a LOFF phase, we relax $%
\mu ^{\prime }=0$, so we have two variational integrals to find the
minimum of $\{\mu ,\mu ^{\prime }\}$. In this case, we also have
three different phases, excluding the normal phase as the following:

\textbf{s-wave}: In this case with $w=0,\psi \neq 0$, from
minimization of $G$ equation and by remembering that in this case
$<J_{x}>=0$ (switch-off) we find the same result as the balanced
case, as the following:
\begin{equation}
(\mu -\mu ^{\prime })_{\min }=9.46653.
\end{equation}%
We used the same quantities for the balanced s-wave phase. The
numerical result as it has been reported in \cite{Amado:2013lia} is
$(\mu -\mu ^{\prime })\approx 8.127$. So, the result is very close
to the numerical estimates.

\textbf{p-wave}: In this case with $\psi =0,w\neq 0$ we set
$D_{+}=0$, from minimization we find that:
\begin{equation}
|\mu ^{\prime }|_{min}=3.71044.
\end{equation}%
We used the minimization technique firstly on the $G$ equation to find
the minimum of the $\mu ^{\prime }$ . The numerical value obtained
is $|\mu ^{\prime }|\geq 3.65$ \cite{Amado:2013lia}. We have a great
agreement between the results. This phase transition corresponds to
the breaking of the $U(1)\times U(1)$ to $U(1)$.

\section{Existence of $s+p$ Phase}

We study the mixed phase $s+p$ separately in this section. Due to
the mixing terms of $\mu ^{\prime },$ $D_{+}$ in F equation and $\mu
-\mu ^{\prime },<J_{x}>$ in the $G$ equation, in the case of the
unbalanced superconductor it is a hard job to find a numeric for
$\mu ^{\prime }$ since we are not aware about the functionality of
the $D_{+},<J_{x}>$ versus the chemical potentials. Before to
studying this numerical estimation we rewrite the system of the $F,G$
equations in the following integral form:
\[
\int_{0}^{1}{dzFG\delta Q^{0}}+\int_{0}^{1}{dzFG\delta (\lambda
W)}+\epsilon
_{1}\int_{0}^{1}{dzQ_{F}^{1}FG^{3}}-\epsilon _{2}\int_{0}^{1}{%
dzQ_{G}^{1}F^{3}G}=0,
\]%
here $\delta f=f_{F}-f_{G}$, and the up script $0$ refers to that
part of the quantity  without the perturbation parameters
$\epsilon _{i}$. since in the numerical study of the s+p phase we
distinguish three regimes depending on the value of the $\delta
=\frac{\mu ^{\prime }}{\mu }$.

\subsection{\textbf{Balanced case}}:
Here $w,\psi \neq 0$, so consequently, $\{<J_{x}>,D_{+}\}\neq 0$, we
will have a minimum but a two variable function $\mu (a,b)$ for a
set of the
trial functions as $\{y_{G},y_{F}\}$. Actually, when we look at the (\ref%
{mu2}), we observe that in the balanced case with $\mu ^{\prime
}=0,$ always there exists a minimum $D_{+}$. The keynote here is
that in the balanced case always $<J_{x}>\sim (\mu -\mu _{c})$.
Indeed in this case we always have
\[
<J_{x}>\sim (\mu -\mu _{c})=\gamma _{1,-1/2}(\mu -\mu _{c}),\ \
\gamma _{1,-1/2}=(0.2157a^{2}|\mu _{c}^{\prime }-\mu _{c}|)^{-1/2}.
\]%
The proportionality coefficient is a constant. Since we know only the approximated expression of $J_ {x} >$ near the criticality and if we apply the minimization we will have a nonlinear expression which cannot be solved analytically, we will obtain the order of the critical chemical potential. We obtain the following expressions for $\mu _{m}=\min\{\mu \} $:
\[
\mu _{m}^{\pm }=\frac{\mu ^{\ast }}{2}\pm \sqrt{\Delta },\ \ \Delta
=\mu _{0}^{2}-\mu ^{\ast }\mu _{c}+\frac{(\mu ^{\ast })^{2}}{4}.
\]%
We have two values of $\mu _{m}$ as the minimum of the $s+p$ phase
chemical potential. It shows that always there exists an
intermediate phase but due to the leakage of the complete form of
the $<J_{x}>$ a more precise estimation is not possible. So,
qualitatively
\[
\mu ^{(s+p)-wave}=\frac{\mu ^{\ast }}{2}\pm \sqrt{\Delta }.
\]

The numeric is more precise than our estimate. In fact, the numerical estimation is still valid for another value of the $J_ {x} >$. The reason of the phase transition is the breaking of rotational symmetry \cite{Amado:2013lia}. 

\subsection{\textbf{Unbalanced configurations}}:
Here both $w,\psi \neq 0$, so consequently $\{J>,D_{+}\}\neq 0$. In this intermediate case we will use both of the F, G equations simultaneously. In a similar method as discussed in the previous section we use the approximation formulas of the condensates near the critically and we obtain:
\[
\Big[\frac{\mu _{m}^{\prime }}{\mu _{m}-\mu _{m}^{\prime }}\Big]^{2}=\frac{%
\mu _{3}\mu _{4}(\mu _{m}^{\prime }-\mu _{c}^{\prime })(\mu _{m}-\mu
_{c}+\mu _{1})}{\mu _{2}\mu _{4}(\mu _{m}-\mu _{c})^{2}+\mu _{2}\mu
_{3}(\mu _{m}^{\prime }-\mu _{c}^{\prime })^{2}+\beta \mu _{2}\mu
_{3}\mu _{4}(\mu _{m}^{\prime }-\mu _{c}^{\prime })}.
\]%
Here $\mu _{i},\beta $ are constants. The general solution for the
cubic equation (\ref{eq3}) can be written as the quadratures. But
physically we can also solve it for the following three distinct
cases:

Case $\mu >>\mu ^{\prime }$ but $\mathcal{O}(\mu ^{\prime })=0$: In
this case we have:
\[
\mu _{m}\sim \mu _{c}-\mu _{1}.
\]%
It gives us the minimum of the $\mu $ independent of the $\mu _{m}^{\prime }$%
.

Case $\mu >>\mu ^{\prime }$ but $\mathcal{O}(\mu ^{\prime 2})=0$: In
this case we have:
\[
\mu _{m}\sim \sqrt[3]{\frac{-\mu _{c}^{\prime }(-\mu _{4}\mu
_{c}^{2}+2\mu _{4}\mu _{c}\mu _{m}-\mu _{4}\mu _{m}^{2}+\mu _{3}\mu
_{c}^{\prime }+\beta \mu _{3}\mu _{4}\mu _{c}^{\prime })}{\mu
_{4}}}.
\]%
It gives us the minimum value of the $\mu $ in terms of the $\mu
_{m}^{\prime }$.

Case $\mu <<\mu ^{\prime }$ . In this case we have:
\[
\mu _{c}^{\prime }\sim \ \frac{\mu _{2}\mu _{4}(\mu _{c}-\mu
_{m})}{\mu _{3}\mu _{2}-\mu _{1}\mu _{4}+\beta \mu _{2}\mu _{4}+\mu
_{4}\mu _{c}-\mu _{4}\mu _{m}}.
\]%
It gives us the minimum of the $\mu _{c}^{\prime }$ in terms of the
$\mu _{i} $.

So, the existence of $s+p$ phase is verified qualitatively by these
solutions.

\section{On relation of condensates and chemical potentials}

In this section we want to find an approximation for the relations $%
\{<J_{x}>,D_{+}\}$ and $\{\mu -\mu ^{\prime },\mu ^{\prime }\}$
qualitatively. Following the series expansions obtained in \cite{Cai:2011ky}%
, and using the asymptotic AdS solutions for $\{w,\psi \}$ we write
down the equations (\ref{eq22}, \ref{eq33}) as perturbations around
the following zero order solutions:

\[
\Phi \approx \mu _{c}+D_{+}\chi (z),\Theta \approx \mu _{c}^{\prime
}+<J_{x}>\zeta (z),
\]%
as the following forms:

\begin{eqnarray}
\chi ^{\prime \prime }-\frac{D_{+}z^{2}G^{2}}{1-z^{3}}(\mu -\mu
^{\prime
})_{c} &=&0,  \label{chi} \\
\zeta ^{\prime \prime }+\frac{z^{2}G^{2}}{1-z^{3}}(\mu -\mu ^{\prime })_{c}%
\frac{D_{+}^{2}}{<J_{x}>}-\frac{z^{2}F^{2}}{1-z^{3}}\mu _{c}^{\prime
} &<&J_{x}>=0.  \label{zeta}
\end{eqnarray}%
With trial functions $y_{G},y_{F}$ the solutions of the equations (\ref{chi}%
, \ref{zeta}) is evaluated at $z=0$. Since
\begin{eqnarray}
\Phi \approx \mu -\rho z\approx \mu _{c}+D_{+}\Big(\chi (0)+\chi
^{\prime }(0)z+\frac{1}{2}\chi ^{\prime \prime 2}+..\Big).
\label{series1}
\end{eqnarray}%
So, we have
\begin{equation}
\mu \approx \mu _{c}+D_{+}\chi (0)
\end{equation}%
and since the general solution for (\ref{chi}) reads as the
following:
\begin{eqnarray}
\chi (z) &=&c_{0}+c_{1}z+D_{+}\Big(-0.05\,{z}^{5}{a}^{2}\mu +0.33\,{z}^{3}{a}%
^{2}\mu -0.5\,{z}^{2}{a}^{2}\mu \\
&&-0.5\,{a}^{2}\mu \,z\ln \left( {z}^{2}+z+1\right) +z\mu \,{a}^{2}-0.25\,{a}%
^{2}\mu \,\ln \left( {z}^{2}+z+1\right)  \nonumber \\
&&+1.7321\,{a}^{2}\mu \,\arctan \left( 1.1547\,z+0.57736\right) z-0.75\,{a}%
^{2}\mu \,\ln \left( \left( 1.1547\,z+0.57736\right) ^{2}+1\right)
\nonumber
\\
&&+0.05\,{z}^{5}{a}^{2}\mu ^{\prime }-0.33\,{z}^{3}{a}^{2}\mu
^{\prime
}+0.5\,{z}^{2}{a}^{2}\mu ^{\prime }+0.5\,{a}^{2}\mu ^{\prime }\,z\ln \left( {%
z}^{2}+z+1\right) -z\mu ^{\prime }\,{a}^{2}  \nonumber \\
&&+0.25\,{a}^{2}\mu ^{\prime }\,\ln \left( {z}^{2}+z+1\right) -1.7321\,{a}%
^{2}\mu ^{\prime }\,\arctan \left( 1.1547\,z+0.57736\right) z  \nonumber \\
&&+0.75\,{a}^{2}\mu ^{\prime }\,\ln \left( \left(
1.1547\,z+0.57736\right) ^{2}+1\right) \Big)\Big).  \nonumber
\label{chisolution}
\end{eqnarray}%
From this solution we obtain:
\begin{equation}
\chi (0)\approx c_{0}+0.2157a^{2}(\mu -\mu ^{\prime })_{c}D_{+}
\end{equation}%
Without any loss of the generality we set $c_{0}=0$. Consequently we
obtain:
\begin{equation}
D_{+}\approx \sqrt{\frac{\mu -\mu _{c}}{0.2157a^{2}(\mu _{c}^{\prime
}-\mu _{c})}}=C\sqrt{\mu -\mu _{c}}.  \label{D+}
\end{equation}%
It shows that the critical exponent $\frac{1}{2}$ is also recovered
in this case.

For the current $<J_{x}>$ we repeat the calculation by integration
of $\zeta$ equation:
\begin{eqnarray}
&&\zeta(z)=c_0+c_1
z-1/3\,C+1/2\,B{z}^{2}{a}^{2}-1/2\,B{a}^{2}\arctan \left(
2/3\,\sqrt {3 }z+1/3\,\sqrt {3} \right) \sqrt {3} \\
&& +1/6\,C{b}^{2}\arctan \left( 2/3\, \sqrt {3}z+1/3\,\sqrt {3}
\right) \sqrt {3}+2/3\,Cb\ln \left( z-1 \right)
z-1/3\,B{z}^{3}{a}^{2}-1/3\,Cbz\ln
\left( {z}^{2}+z+1 \right)  \nonumber \\
&& +1/6\,C{b}^{2}z\ln \left( {z}^{2}+z+1 \right) +1/3\,Cb \arctan
\left(
2/3\,\sqrt {3}z+1/3\,\sqrt {3} \right) \sqrt {3}+1/2\,B {a}^{2}z\ln \left( {z%
}^{2}+z+1 \right)  \nonumber \\
&& +2/3\,Cb\arctan \left( 2/3\, \sqrt {3}z+1/3\,\sqrt {3} \right) \sqrt {3}%
z-1/2\,C{b}^{2}{z}^{2}+1/20 \,B{z}^{5}{a}^{2}-1/20\,{z}^{5}C{b}^{2}
\nonumber \\
&& +1/3\,{z}^{3}Cb-1/3\,C\sqrt {3} \arctan \left( 1/3\, \left(
2\,z+1
\right) \sqrt {3} \right) +3/4\,B{a }^{2}\ln \left( \left( 2/3\,\sqrt {3}%
z+1/3\,\sqrt {3} \right) ^{2}+1 \right)  \nonumber \\
&& -1/4\,C{b}^{2}\ln \left( \left( 2/3\,\sqrt {3}z+1/3\,\sqrt {3}
\right)
^{2}+1 \right) -1/3\,C\ln \left( z-1 \right) z-1/3\,Cz \ln \left( {z}%
^{2}+z+1 \right)  \nonumber \\
&& -1/2\,Cb\ln \left( \left( 2/3\, \sqrt {3}z+1/3\,\sqrt {3} \right)
^{2}+1 \right) -B{a}^{2}\arctan \left( 2/3\,\sqrt {3}z+1/3\,\sqrt
{3} \right)
\sqrt {3}z  \nonumber \\
&& +1/3\,C{b}^{2 }\arctan \left( 2/3\,\sqrt {3}z+1/3\,\sqrt {3}
\right)
\sqrt {3}z-1/3 \,C{b}^{2}\ln \left( z-1 \right) z  \nonumber \\
&& +2/3\,Cb-1/3\,C{b}^{2}-1/3\,\sqrt { 3}\arctan \left( 1/3\, \left(
2\,z+1 \right) \sqrt {3} \right) Cb+1/2 \,\sqrt {3}\arctan \left(
1/3\, \left(
2\,z+1 \right) \sqrt {3} \right) B{a}^{2}  \nonumber \\
&& +1/6\,\sqrt {3}\arctan \left( 1/3\, \left( 2\,z+1 \right) \sqrt
{3}
\right) C{b}^{2}-zB{a}^{2}+1/4\,\ln \left( {z}^{2} +z+1 \right) B{a}%
^{2}-1/6\,\ln \left( {z}^{2}+z+1 \right) Cb  \nonumber \\
&& +1/12\, \ln \left( {z}^{2}+z+1 \right) C{b}^{2}-2/3\,C\ln \left(
z-1
\right) b+1/3\,C\ln \left( z-1 \right) {b}^{2}  \nonumber \\
&& -1/6\,\ln \left( {z}^ {2}+z+1 \right) C+1/3\,C\ln \left( z-1
\right) +zC \nonumber
\end{eqnarray}
Here $B=\frac{a^2(\mu_c-\mu^{\prime
}_c)(\mu-\mu_c)}{<J_x>},C=\mu^{\prime }_c<J_x>$. By evaluating the
solution at $z=0$ we find:
\begin{eqnarray}
\zeta (0)=c_0+\left( - 2.0944\,b+ 1.0472\,{b}^{2}+ 0.52284\,b- 0.10294\,{b}%
^{2}- 0.63564+ 1.0472\, \right) C+ 0.21574\,B{a}^{2}.
\end{eqnarray}%
Since $B\gg1,C\ll1$ so ,
\begin{equation}
\mu ^{\prime }\approx \mu _{c}^{\prime }+<J_{x}>\zeta (0)
\end{equation}%
after a simple algebraic manipulation we find:
\begin{eqnarray}
<J_x>\approx {\frac { 24{C}^{2}{a}^{2}\left( \mu^{\prime }_c- \mu_c
\right)
\left( \mu-\mathit{\mu_c } \right) }{\sqrt {\mu^{\prime }-\mathit{%
\mu^{\prime }_c}} (\mu^{\prime }_c)^{1/2}\, \left( - 78578\,b+ 47213\,{b}%
^{2}+ 20578 \right)^{1/2} }} \\
+ \,{\frac{223.61 \sqrt {\mu^{\prime }-\mathit{\mu^{\prime }_c}}}{\mathit{%
(\mu^{\prime }_c)^{1/2}}\, \left( - 78578\,b+ 47213\,{\ b}^{2}+
20578 \right)^{1/2} }} .  \nonumber  \label{J}
\end{eqnarray}%
It is a completely different expression and it shows that in the s+p
phase
why the vector current is a function of the all chemical potentials $%
\mu,\mu^{\prime }$.

Now we can compute the relation between the chemical potentials and
the charge densities. Using the series expansion given in
(\ref{series1}) we obtain:
\[
\rho =-D_{+}\chi ^{\prime }(0).
\]%
The same relation is valid, but with simple replacement of $%
<J_{x}>\rightleftharpoons D_{+},\rho ^{\prime }\rightleftharpoons
\rho ,\zeta \rightleftharpoons \chi $. Using (\ref{chisolution}) we
find:
\[
\chi ^{\prime }(0)=c_{0}+0.9069a^{2}D_{+}(\mu -\mu ^{\prime })_{c}.
\]%
So,
\[
\rho =-D_{+}c_{0}-0.9069a^{2}D_{+}^{2}(\mu -\mu ^{\prime })_{c}.
\]%
Using (\ref{D+}) we finally have:
\[
\rho =-c_{0}C(\mu -\mu _{c})^{1/2}-0.9069a^{2}C^{2}(\mu -\mu
^{\prime })_{c}(\mu -\mu _{c})\equiv \beta _{1/2}(\mu -\mu
_{c})^{1/2}+\beta _{1}(\mu -\mu _{c}).
\]%
Here $\beta _{i}$ denotes the coefficient of $(\mu -\mu _{c})^{i}$. When $%
c_{0}=0$ we have:
\[
\rho \approx \beta _{1}(\mu -\mu _{c}).
\]%
The linear well known approximation $\rho =(\mu -\mu _{c})$ still
now is valid as a linear approximation for $c_{0}=0$ or when $\mu
>>\mu _{c}$. But due to the mixing of the s-, p-phases the behavior
of the $\rho -(\mu ,\mu ^{\prime })$ does not remain linear due to
the term $-c_{0}C(\mu -\mu _{c})^{1/2}$ . As we conclude that in the
mixed phase when the system is far from the critical point the
charge density of the mixed phase is larger than the pure phases of
s or p. So, the free energy of the mixed phase $s+p$ is smaller than
the pure single phases s or p. It means $F_{s+p}<F_{s,p},$ as it has
been shown in \cite{Amado:2013lia}, the free energy is a linear
function of the $-\rho ,-\rho ^{\prime },$ and also it has a non
local integral term as the following:
\[
F=-\frac{1}{2}\Sigma _{i}\mu _{i}\rho _{i}+\int {\ dr\{\mathbf{fields}+%
\mathbf{fields}^{\prime }+..\}}.
\]

Since the system thermodynamical goes to the minimum states with
minimum of F, so in the competition between these phases, $s+p$ has
a greater chance to be the dominant phase and it is the preferred
phase. In this duel between the phases, the system prefers to remain
in a mixed phase (s+p). The same conclusion is reported in
\cite{Amado:2013lia} by numerically studying the behavior of the
equation F. We pertinently mention here that since the analytical
estimations of $\mu _{c},\mu _{c}^{\prime }$ are very close to the
numerical ones so the evolutionary scheme of the system, after these
critical chemical potentials, will act as a second order phase
transition.

Now we compute the relation of $\rho ^{\prime },(\mu ^{\prime }-\mu
_{c}^{\prime })$ following the same strategy. We have:
\[
\zeta ^{\prime }(0)=<J_{x}>\left(
-1.0472-0.74489\,{b}^{2}+2.6990\,b\right)
\mu _{{c}}^{\prime }-\,c_{{1}}-0.90695{c}^{2}{a}^{2}\,{\frac{\left( \mu _{{c}%
}-\mu _{{c}}^{\prime }\right) \left( \mu -\mu _{{c}}\right)
}{<J_{x}>}}.
\]%
So,
\begin{eqnarray}
\rho ^{\prime } &=&<J_{x}>^{2}\left( 1.0472+0.74489\,{b}^{2}-2.6990\,b%
\right) \mu _{{c}}^{\prime }+<J_{x}>\,c_{{1}} \\
&&+0.90695\left( \mu _{{c}}-\mu _{{c}}^{\prime }\right)
{c}^{2}\left( \mu -\mu _{{c}}\right) {a}^{2}.  \nonumber
\end{eqnarray}%
Since,
\[
<J_{x}>=\frac{\hat{A}(\mu -\mu _{c})}{\sqrt{\mu ^{\prime }-\mu _{c}^{\prime }%
}}+\hat{B}\sqrt{\mu ^{\prime }-\mu _{c}^{\prime }},
\]%
here $\hat{A},\hat{B}$ are two proportionality constants, so near
the critical point when $\mu \rightarrow \mu _{c},\mu ^{\prime
}\rightarrow \mu _{c}^{\prime }$ we have:
\begin{eqnarray}
\rho ^{\prime } &=&\alpha _{2}^{-1}(\mu -\mu _{c})^{2}(\mu ^{\prime
}-\mu _{c}^{\prime })^{-1}+\alpha _{1}^{0}(\mu -\mu _{c})+\alpha
_{1}^{1}(\mu
^{\prime }-\mu _{c}^{\prime })(\mu -\mu _{c}) \\
&&+\alpha _{1}^{-1/2}\frac{(\mu -\mu _{c})}{\sqrt{\mu ^{\prime }-\mu
_{c}^{\prime }}}+\alpha _{0}^{1/2}\sqrt{\mu ^{\prime }-\mu
_{c}^{\prime }}, \nonumber
\end{eqnarray}%
here $\alpha _{i}^{j}$ denotes the coefficient of the term $(\mu
-\mu _{c})^{i}(\mu ^{\prime }-\mu _{c}^{\prime })^{j}$. So far, also
the linear approximation exists as the limiting case while the other
non-linear terms removed.

So, these generalized expressions show that why a mixed critical phase of $%
(s+p)$ exists in a system of  the $U(2)$ fields under symmetry breaking to
$U(1)$.

\section{Summary}

In this paper, for the first time in the literature, we have studied
the analytical mixed phase transitions from normal phase to an
intermediate phase (s+p) as well as a p-wave in a newly proposed model
of holographic superconductors. By using the variational method, we
have shown that in balanced and unbalanced configurations, a minimum
of critical chemical potential exists in which the system undergoes
to the s-wave or p-wave and also in an intermediate s+p phases, due
to the spontaneous symmetry breaking into the U(1) symmetry. Also,
we have found the condensation values as functions of the dual
chemical potentials analytically. Our results have a
reasonably good qualitative agreement with the numerical results \cite%
{Amado:2013lia}. \vspace{12pt}

\section*{Acknowledgments}

We want to thank Daniele Musso, Zhang-Yu Nie, A.~Amoretti,
A.~Braggio, N.~Maggiore and N.~Magnoli for useful comments and also the anonymous reviewer for enlightening comments related to this work.






\end{document}